\documentclass{PoS}
\pdfoutput=1

\newcommand{\bea}{\begin{eqnarray}}
\newcommand{\eea}{\end{eqnarray}}
\newcommand{\beq}{\begin{eqnarray}}
\newcommand{\eeq}{\end{eqnarray}}

\newcommand{\VA}{\langle A^2 \rangle}
\newcommand{\eq}[1]{Eq.~(\ref{#1})}

\title{Dynamical quarks effects on the gluon propagation and chiral symmetry restoration}

\ShortTitle{Dynamical quarks effects on the gluon propagation}

\author{A. Bashir\\
       Universidad Michoacana de San Nicol\'as
de Hidalgo, Edificio C-3, Ciudad Universitaria, Morelia, Michoac\'an 58040, M\'exico. \\
        E-mail: \email{adnan@ifm.umich.mx}}
\author{A. Raya\\
       Universidad Michoacana de San Nicol\'as
de Hidalgo, Edificio C-3, Ciudad Universitaria, Morelia, Michoac\'an 58040, M\'exico. \\
\& \\
Facultad de F\'{\i}sica, Pontificial Universidad Cat\'olica de Chile, Casilla 306, Santiago 22, Chile\\ 
        E-mail: \email{raya@ifm.umich.mx}}
\author{\speaker{J. Rodr\'{\i}guez-Quintero}\\
        Departamento de F\'{\i}sica Aplicada, Facultad de Ciencias
Experimentales, Universidad de Huelva, Huelva 21071, Spain.\\
\& \\
CAFPE, Universidad de Granada, E-18071, Granda, Spain. \\  
        E-mail: \email{jose.rodriguez@dfaie.uhu.es}}

\abstract{We exploit the recent lattice results for the infrared gluon
propagator with light dynamical quarks and solve the gap equation
for the quark propagator. Chiral symmetry breaking 
and confinement (intimately tied with the analytic properties of 
QCD Schwinger functions) order parameters are then studied.}

\FullConference{QCD-TNT-III-From quarks and gluons to hadronic matter: A bridge too far?,\\
		2-6 September, 2013\\
		European Centre for Theoretical Studies in Nuclear Physics and Related Areas (ECT*), Villazzano, Trento (Italy)}

\begin{document}

\section{Introduction}

This contribution is mainly devoted to report on the results of Ref.~\cite{Bashir:2013zha}, 
where we dealt with the dynamical quarks effects on the gluon propagation and applied this knowledge to 
the analysis of the chiral symmetry restoration with increasing number of light fermion flavours. 
Quantum chromodynamics (QCD) with large number of massless fermion
flavors has seen a resurgence of interest due to its connection
with technicolor models, originally proposed by Weinberg and
Susskind~\cite{Weinberg:1979}, which fall into the category of
\emph{Beyond the Standard Model Theories}. They possess
intrinsically attractive features. They do not resort to
fundamental scalars to reconcile local gauge symmetry with massive
mediators of interaction and have close resemblance with
well-studied fundamental strong interactions, i.e., QCD. However,
their simple versions do not live up to the experimental
electroweak precision constraints, in particular the ones related
to flavor changing neutral currents. Walking models containing a
conformal window and an infrared fixed point can possibly cure
this defect and become phenomenologically
viable~\cite{Holdom:1985}. This scenario motivates the
investigation of QCD for similar characteristics. One looks for
such behavior of QCD for large number of light flavors
\emph{albeit} less than the critical value where asymptotic
freedom sets in, i.e., $N_f^{c_1} = 16.5$, a Nobel prize winning
result known since the advent of QCD,~\cite{Wilczek:1973}. Just as
$N_f$ dictates the peculiar behavior of QCD in the ultraviolet, we
expect it to determine the onslaught of its emerging phenomena in
the infrared, i.e., chiral symmetry breaking and confinement.

Whereas the self interaction of gluons provides anti-screening,
the production of virtual quark-antiquark pairs screens and
debilitates the strength of this interaction of non abelian
origin. For real QCD, light flavors are small in number and hence
yield to the gluonic influence which triggers confinement and
chiral symmetry breaking. One needs to establish if there is
another critical value $N_f^{c_2} < N_f^{c_1}$ which can
sufficiently dilute the gluon-gluon interactions to restore chiral
symmetry and deconfine color degrees of freedom. Such a phase
transition lies at the non perturbative boundary of the
interactions under scrutiny and hence we cannot expect to extract
sufficiently reliable information from  multiloop calculations of
the QCD $\beta$-function. Purely non perturbative techniques are
required to tackle the problem. Lattice studies in the infrared
indicate that just below $N_f^{c_1}$, chiral symmetry remains
unbroken and color degrees of freedom are
unconfined~\cite{Appelquist:2009}. Below this conformal window,
for an $8< N_f^{c_2} < 12$, the evolution of the beta function in
the infrared is such that QCD enters the phase of dynamical mass
generation as well as confinement.

Modern lattice analyses for this matter appear to strongly argue in 
favour of a restoration for the chiral symmetric phase taking place 
somewhere between $N_f \sim 8$ and $N_f \sim 10$~\cite{Iwasaki:2012,Aoki:2013}. 
In particular, the authors of ref.~\cite{Aoki:2013}, with their study 
of the meson spectrum in lattice QCD with eight light flavours using 
the Highly Improved Staggered Quark action, gathered some striking 
evidences that $N_f=8$ QCD still lies in the broken chiral symmetry 
phase but, at the sime time, suffering the effects from a remnant of 
the infrared conformality (a large anomalous dimension for the 
quark mass renormalization constant) indicating that the unbroken 
phase is recovered near above $N_f \sim 8$. In the present work, 
we intend to combine the Schwinger-Dyson machinery, well adjusted 
to account for QCD phenomenology in the pion sector, with the very 
last lattice data including twisted-mass dynamical light flavours 
in order to provide with a model for the chiral restoration 
mechanism, in quantitative agreement with the above mentioned 
lattice studies.  

\section{Chiral phase transition picture from Schwinger-Dyson 
and Lattice gluon propagators}

In continuum, Schwinger-Dyson equations (SDEs) of QCD provide an
ideal framework to study its infrared properties,~\cite{SD:1949}.
These are the fundamental equations of any quantum field theory,
linking all its defining Green functions to each other through
intricately coupled nonlinear integral equations. As their formal
derivation through variational principle makes no appeal to the
weakness of the interaction strength, they naturally connect the
perturbative ultraviolet physics with its emerging non
perturbative properties in the infrared sector within the same
framework. The simplest two-point quark propagator is a basic
object to analyze dynamical chiral symmetry breaking and
confinement. Within the formalism of the SDEs, the inverse quark
propagator can be expressed as 
\bea\label{eq:gap}
 S^{-1}(p) = {\cal Z}_2 (i \gamma \cdot p + m) +  \Sigma(p) \ ,
 \eeq
where $\Sigma(p)$ is the quark self energy:
 \bea\label{eq:sigma}
   \Sigma(p)= {\cal Z}_1 \int \frac{d^4q}{(2 \pi)^4} \, g^2 \Delta_{\mu
   \nu}(p-q) \frac{\lambda^a}{2} \gamma_{\mu} S(q)
   \Gamma_{\nu}^a(q,p) \;,
 \eea
where ${\cal Z}_1={\cal Z}_1(\mu^2,\Lambda^2)$ and ${\cal
Z}_2={\cal Z}_2(\mu^2,\Lambda^2)$ are the renormalization
constants associated respectively with the quark-gluon vertex and
the quark propagator. $\Lambda$ is the ultraviolet regulator and
$\mu$ is the renormalization point. The solution to this equation
is
\bea\label{eq:Sm1}
  S^{-1}(p) = \frac{i \gamma \cdot p + M(p^2)}{Z(p^2,\mu^2)} \ ,
\eeq
where $Z(p^2,\mu^2)$ is the quark wavefunction renormalization and
the quark mass function $M(p^2)$ is renormalization group
invariant. This equation involves the quark-gluon vertex
$\Gamma_{\nu}^a(q,p)$ and the gluon propagator $\Delta_{\mu
\nu}(p-q)$. Very much attention will be paid in the next subsection to the two-point function as 
a crucial input to study the quark propagator. Here, in the following, the only other 
ingredient, the three-point
quark-gluon vertex $\Gamma_{\nu}^a(q,p)$, will be briefly discussed. 

Significant advances have been made in pinning it down through its key attributes in
the ultraviolet and infrared domains~\cite{Vertex:All}. More
recently, the seeds of the most general ansatz for the
fermion-boson vertex appeared in~\cite{Sanchez:2011} and its full
blown extension was presented in~\cite{Bermudez:2012}. 
Furthermore, given the general
nature of constraints and the simplicity of the construction, a
straightforward extension of this approach is expected to yield an
ansatz adequate to the task of representing the
dressed-quark-gluon vertex. Before this is achieved, we restrict
ourselves to an effective though efficacious approach. Following
the lead of Maris {\em et. al.}~\cite{Maris:1998}, we employ the
following suitable ansatz which has sufficient integrated strength
in the infrared to achieve dynamical mass generation:
  \bea
  {\cal Z}_1 g^2 \Delta_{\mu \nu}(p-q) \Gamma_{\nu}(p,q) \rightarrow g^2_{\rm eff}(q^2) \; \Delta^N_{\mu \nu}(p-q,N_f)
  \frac{\lambda^a}{2}
  \gamma_{\nu}\;,
 \eea
 where
 \bea
  \Delta^N_{\mu \nu,N_f}(q) &=& \frac{D(q^2,N_f)}{q^2} \; \left[ \delta_{\mu \nu} - \frac{q_{\mu} q_{\nu}}{q^2}
  \right] \; .
 \eea
In the next subsection, the nonperturbative gluon dressing function, $D(q^2,N_f)$, will appear 
modelled and accounting thus for the effects from the dynamical light flavours on the gluon propagation. 
Concerning the effective coupling, $g_{\rm eff}$, it has been chosen to be
\bea\label{eq:MarisTandy}
g^2_{\rm eff}(q^2) D(q^2,N_f) \ = \ \frac{D(q^2,N_f)}{D(q^2,2)} \ 
g^2_{\rm MT}(q^2) \ ,
\eeq
where $g^2_{MT}$ stands for the effective coupling ansatz proposed by 
Maris {\em et. al.} in ref.~\cite{Maris:1998} and built there to incorporate also the 
nonperturbative information from the gluon propagator (as it is assumed to multiply 
the tree-level one) . Thus, \eq{eq:MarisTandy} for two degenerate light 
fermion flavours allows to reproduce correctly the static as well as dynamic properties 
of mesons below 1 GeV (see for example review~\cite{Review:2012} and references therein)
while, by fixing the right number of flavours in the perturbative tail of $g^2_{MT}$, 
it also reproduces the appropriate ultraviolet behaviour, for any flavour number.

\subsection{Modelling the flavour behaviour for the gluon propagator}

The second ingredient needed to solve Eq.~(\ref{eq:gap}) with (\ref{eq:sigma},\ref{eq:Sm1}) is 
the two-point gluon Green function,  $\Delta_{\mu \nu}$, that has been the object of a 
patient effort, spanning several decades, addressed to
unravel its infrared behaviour. Lattice as well as SDE studies have finally converged on its
massive or so called decoupling solution (see for
example~\cite{Gluon:2009}). After the gluon propagator
solution in the quenched approximation has been chiselled, we now
have the first quantitatively reliable glimpses of its quark
flavor dependence by incorporating $N_f=0,2$ light dynamical quark
flavors~\footnote{The dynamical flavors have been generated,
within the framework of ETM
collaboration~\cite{Baron:2010bv,Baron:2011sf,Blossier:2010ky,Blossier:2011tf},
with the mass-twisted lattice action~\cite{Frezzotti:2000nk},
while $N_f=0$ data have been borrowed
from~\cite{Bogolubsky:2007ud}.} and 2+1+1 (2 light degenerate
quarks, with masses ranging from 20 to 50 [MeV], and two non
degenerate flavors for the strange and the charm quarks, with
their respective masses set to 95 [MeV] and 1.51
[GeV])~\cite{Ayala:2012}. As we demonstrate shortly, in this last
2+1+1 case, the number of light quarks corresponds effectively to
3. This is exactly the result derived from the recently developed
{\it "partially unquenched"} approach to incorporate flavor
effects in the gluon SDE,~\cite{Aguilar:2012rz}. Their work is in
agreement with one of ~\cite{Ayala:2012} when the charm flavor is
assumed to decouple from gluons. It should not be worthless to remark 
that the same $N_f$=2+1+1 gauge fields have been also used to 
compute the running strong coupling and estimate accurately the value 
of $\Lambda_{\overline{\rm MS}}$, in very good agreement with 
experiments~\cite{Blossier:2012ef}.

\begin{figure}[h] 
\begin{center}
\begin{tabular}{cc}
\includegraphics[width=7.5cm]{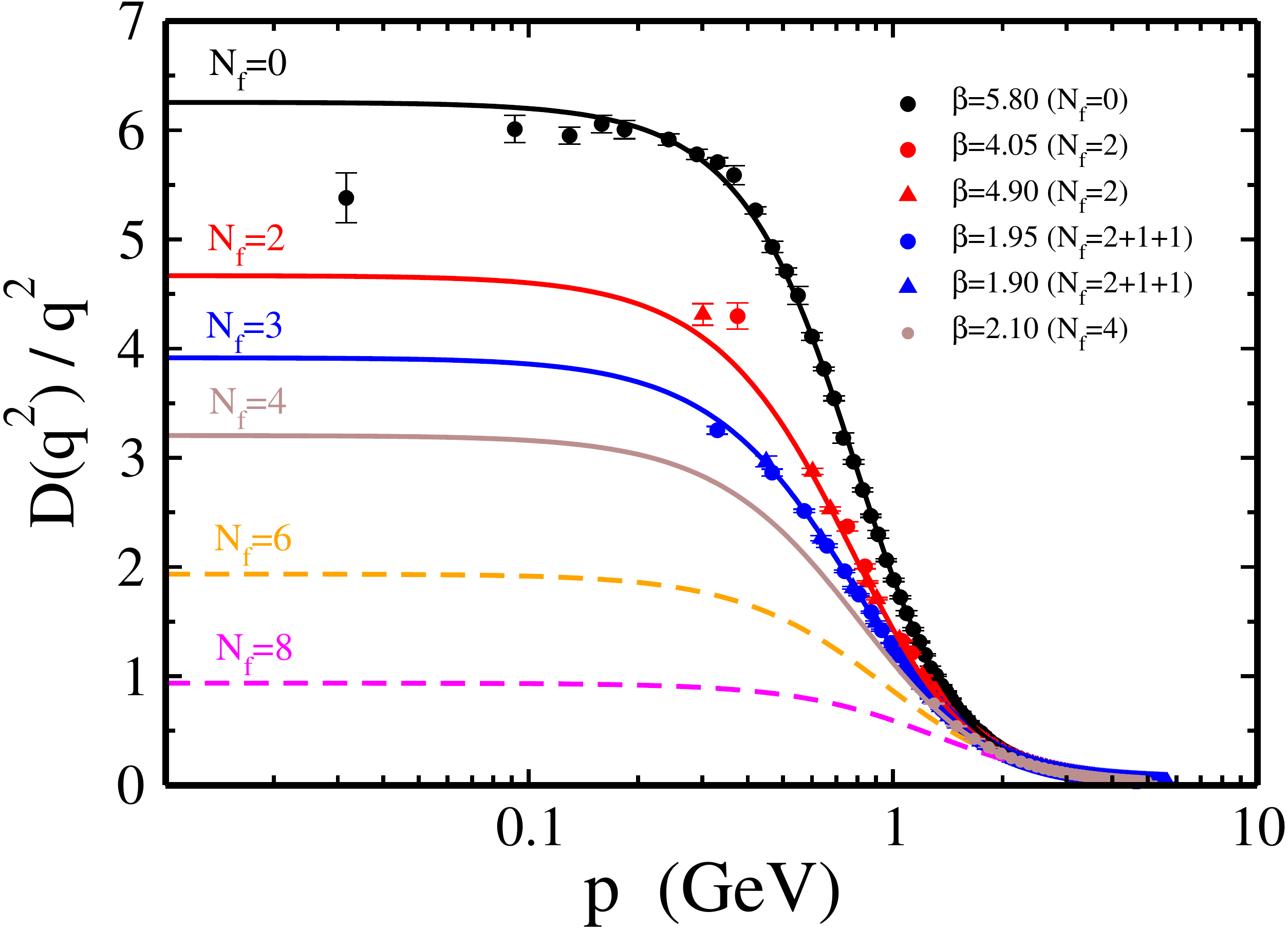} & 
\includegraphics[width=7.5cm]{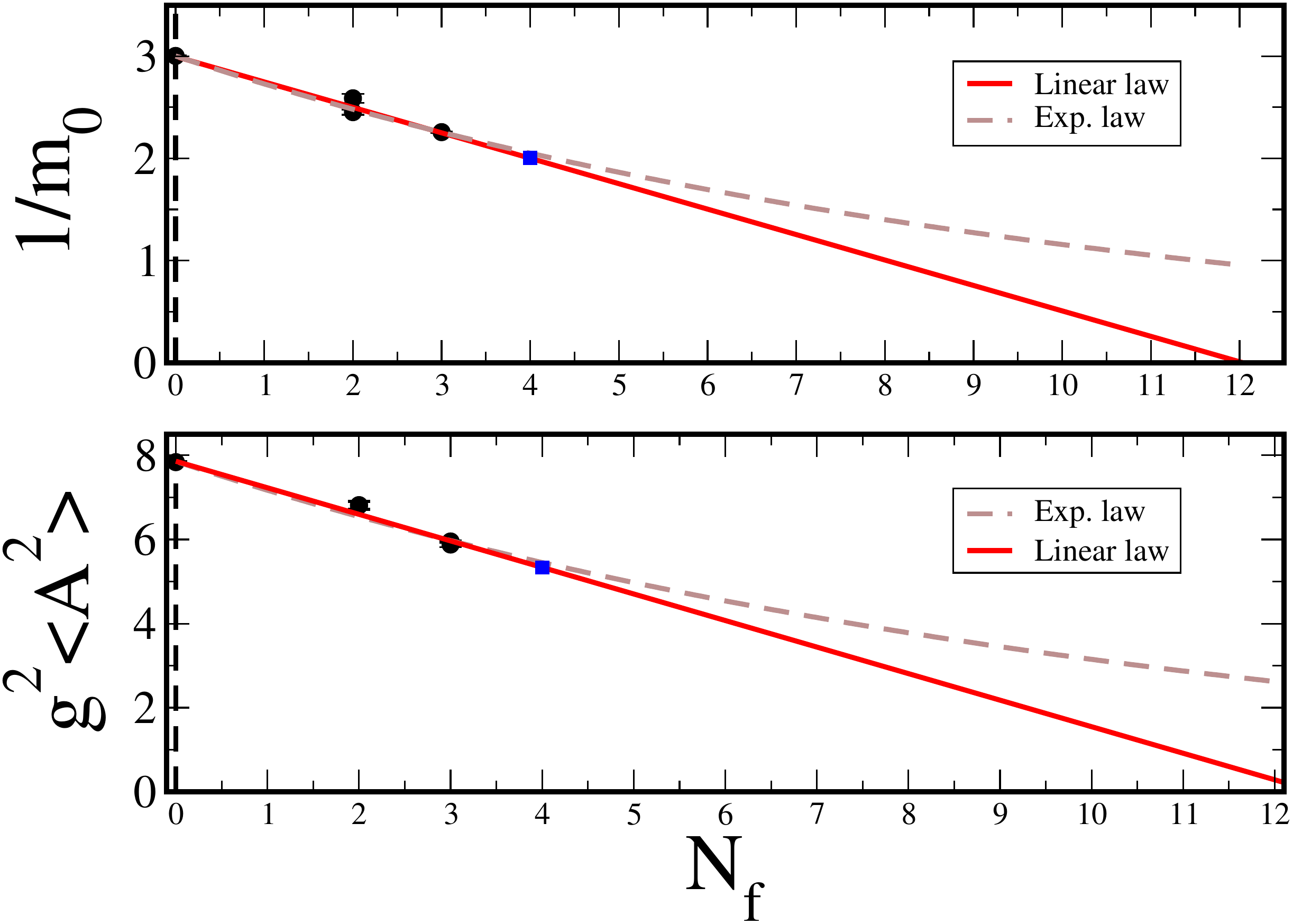}
\end{tabular}
\end{center}
\caption{(Left) Lattice gluon propagator data in terms of momenta for
different number of fermion flavors and fits with  Eq.~(2.7) 
and the parameters of Eqs.~(2.9,2.10). 
(Right) Parameters $g^2<A^2>$ and $1/m_0^2$ in terms of the
numbers of flavors and the fits with Eqs.~(2.9,2.10). 
The blue squares stand for the extrapolated results at $N_f=$4 we used for Fig.~2.
}
\label{fig:parameters}
\end{figure}

\begin{figure}[h] 
{\centering
{\includegraphics[width=12cm]{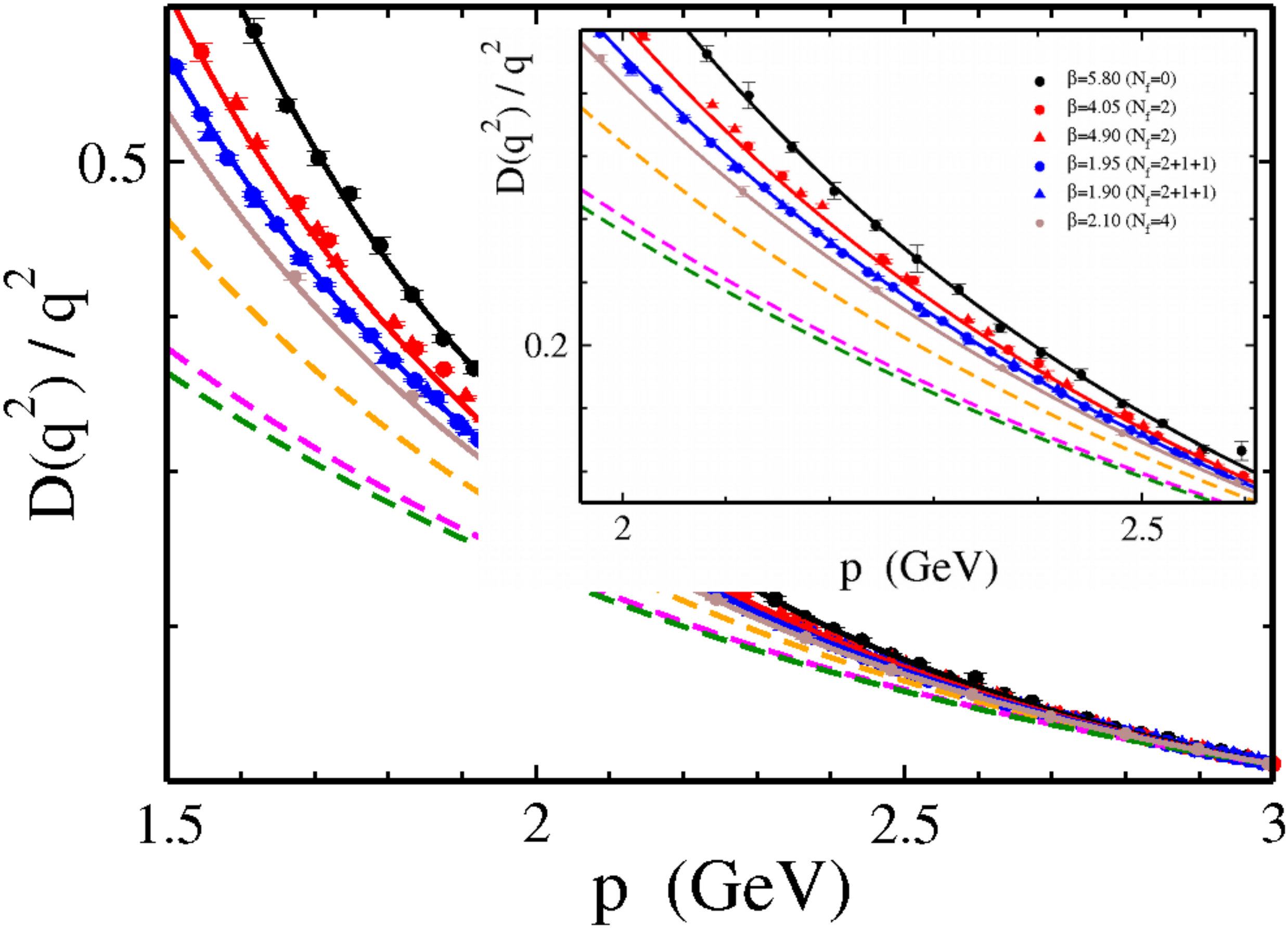}}
\par}
\caption{The same of the left plot of Fig.~1 
but only with the parameters for the linear case and incorporating new small-volume lattice data for 4 degenerate fermion
flavors.}
\label{fig:gluonprop2}
\end{figure}

Moreover, our modern understanding of the flavor dependence of the
gluon propagator provides us with the solid basis to use the
following non perturbative model~\cite{Dudal:2012zx}:
 \beq\label{eq:gluonR} D(q^2) \ = \ \frac{z(\mu^2) \ q^2
 (q^2+M^2)}{\displaystyle q^4 + q^2 \left(M^2-{13} g^2 \VA /24
 \right) + M^2 m_0^2}
 \eeq
to describe the gluon dressing renormalized in MOM scheme at
$q^2=\mu^2$. This model is based on the tree-level gluon
propagator obtained with the Renormalized Gribov-Zwanziger (RGZ)
action~\cite{Dudal:2008sp} which have been shown to describe
properly the lattice data in the infrared sector (see
refs.~\cite{Dudal:2010tf,Dudal:2012zx}). The overall factor
$z(\mu^2)$ is introduced to guarantee the multiplicative MOM
renormalization prescription, namely, $D(\mu^2) = 1$, and implies
no physical consequence as the effective coupling, $g_{\rm eff}$,
is further adjusted to reproduce properly the meson phenomenology.
Then, we apply \eq{eq:gluonR} to reproduce the gluon propagator lattice 
data analyzed in Ref.~\cite{Ayala:2012} and thus fit its mass parameters. 
$M^2$ is related to the condensate of
auxiliary fields, emerging merely to preserve locality for the RGZ
action. A free fit of the lattice data suggests that it does not
depend on the number of fermion flavors (we find $M^2=4.85$
[GeV$^2$]). Dimension two gluon condensate
$\VA$,~\cite{Boucaud:2000nd}, and 
\bea
m_0^2 \ = z(\mu^2) \lim_{q^2\to
0} \frac{q^2}{D(q^2)}
\eeq
are flavor dependent and we look for their best
fits. In order to cover a wide range of possibilities within
reason, we assume their evolution with the flavor number to be
driven either by a simple linear scaling law
\beq\label{eq:masses}
 m_0^{-1}(N_f) &=& {m_0^{-1}(0)} \ (1 - A N_f)  \nonumber    \\
g^2 \VA(N_f) &=& g^2\VA(0) \ (1 - B N_f)  \ ,
\eeq
as data appear to suggest, or by an exponential law
\beq\label{eq:massesexp}
 {m_0^{-1}(N_f)} &=& {m_0^{-1}(0)} \ e^{-A N_f}  \nonumber \\
g^2 \VA(N_f) &=& g^2\VA(0)  \ e^{-B N_f} \ ,
\eeq
which allows for the possibility that the gluon propagator becomes
infinitely massive only when the number of light quark flavors
tends to infinity. The best-fit of the $m_0$ and $g^2\VA$ from
lattice data will require $m_0(0)=0.333$ GeV and
$g^2\VA(0)=7.856$, in both cases, $A=0.083$ and $B=0.080$, for the
linear case, and $A=0.095$ and $B=0.091$, for the exponential one.
Eq.~(\ref{eq:gluonR}) now provides prediction for the gluon
propagator for arbitrarily large $N_f$, as can be seen
in the right plot Fig.~\ref{fig:parameters}, while the right one shows
the corresponding gluon propagator along with the lattice data
superimposed~\cite{Ayala:2012}. We also include some very recent
gluon propagator data obtained from lattice simulations with four
degenerate light twisted-mass flavors~\footnote{The gluon
propagator lattice data for 4 light flavors have been borrowed
from ETMC~\cite{Boucaud:2013bga}. Simulated at small volumes, they are
only available for momenta above 1.25 GeV and hardly allow for a
fit with \eq{eq:gluonR}. Nevertheless, they can be used to check
our modelling of the flavor evolution.}. These new data are rather
well described by \eq{eq:gluonR} evaluated for the mass parameters
extrapolated to $N_f$=4 with \eq{eq:masses} (see the zoomed plot
in Fig.~\ref{fig:gluonprop2}). This observation strongly supports
that $N_f=$2+1+1 gluon data indeed correspond to three light
flavors.

Thus, we can efficaciously model the dilution of the gluon-gluon
interactions with increasing flavor number in order to study the
chiral restoration mechanism. We can now employ the gap equation
to provide quantitative details of chiral symmetry breaking in
terms of the quark mass function for an increasing number of light
quarks.

\subsection{Results}

In the following, we mostly discuss the results obtained by
employing the linear law and state the effect of exponential
extrapolation afterwards. 
Note that we have not considered the flavor dependence which would
arise from the quark-gluon vertex (no explicit handle on this
dependence is available at the moment). Otherwise said, 
as can be seen from \eq{eq:MarisTandy}, we take the effective
coupling $g_{\rm eff}$, in the IR, to depend on $N_f$ only through 
the gluon dressing function. The latter can be fairly justified by the 
results of~\cite{Ayala:2012} (see Eq.(5.2)) which
suggest that an effective coupling can be constructed such that
there is an absence of any flavor dependence in the infrared
region, more precisely starting from $q^2 \lesssim 1$ GeV$^2$.

\begin{figure}[h] 
{\centering {\includegraphics[width=12.5cm]{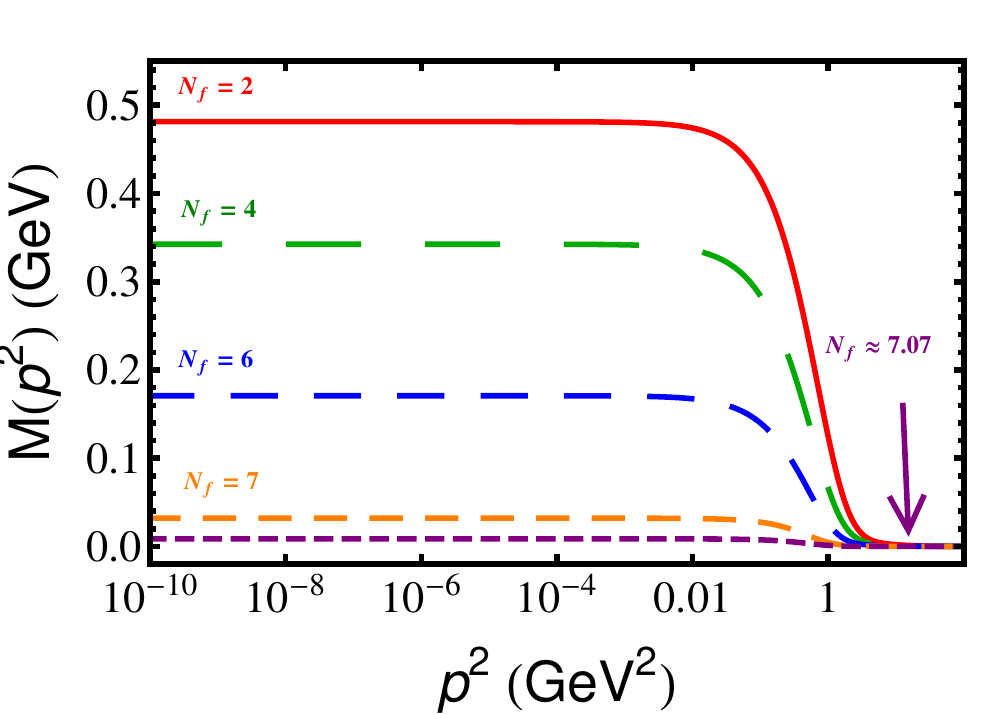}}
\par}
\caption{The quark mass function diminishes height for increasing
light quark flavors (here with Eq.~(2.9) 
Above $N_f \approx 7.07$, only the chirally
symmetric 
solution exists.}\label{fig:quarkmass}
\end{figure}

\begin{figure}[h] 
{\centering {\includegraphics[width=12.5cm]{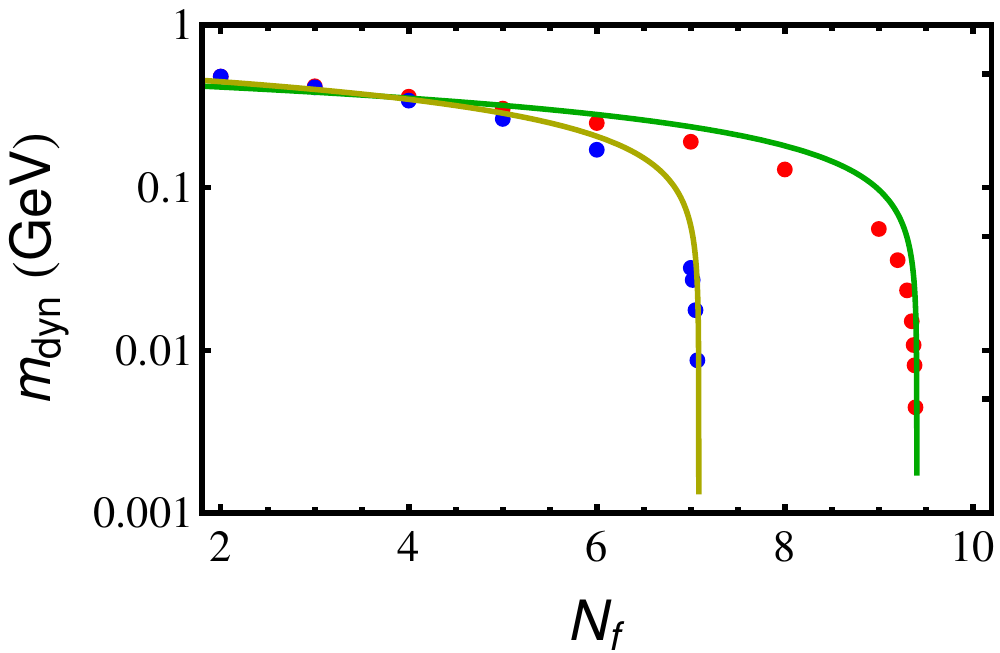}}
\par}
\caption{Quark pole mass in the Euclidean space clearly
demonstrates that chiral symmetry is restored above a critical
number of quark flavors. Blue (red) points correspond to linear (exponential) case. 
The solid line is the mean-field scaling, Eq.(2.11) 
} \label{fig:polemass}
\end{figure}

\begin{figure}[h] 
\begin{center}
\begin{tabular}{cc}
\includegraphics[width=7.5cm,height=6cm]{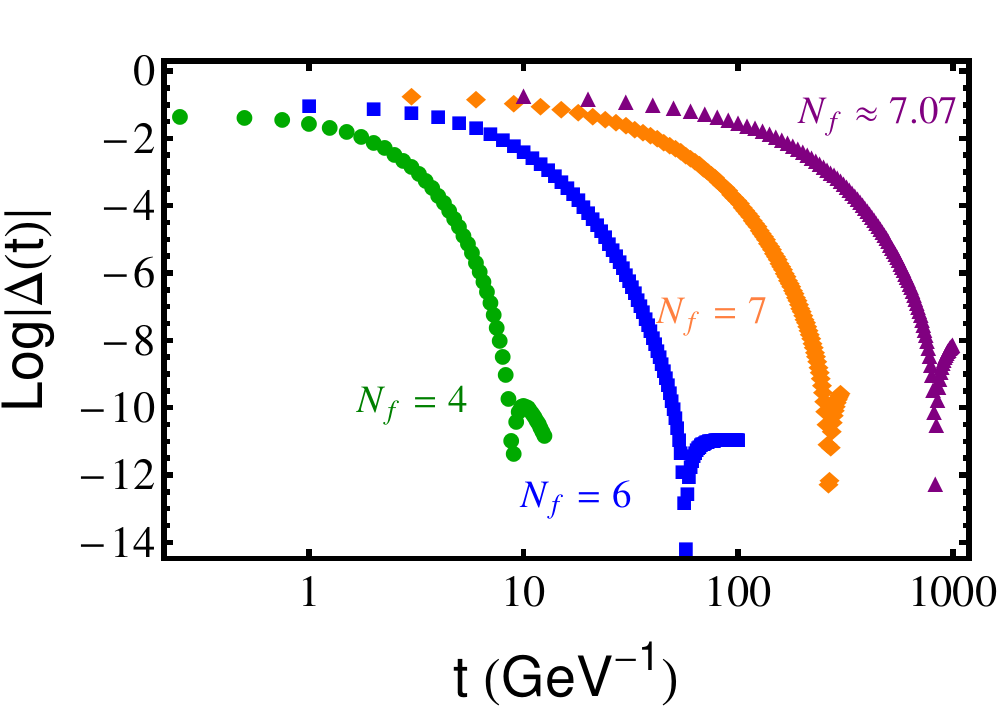} 
&
\includegraphics[width=7.5cm,height=5.5cm]{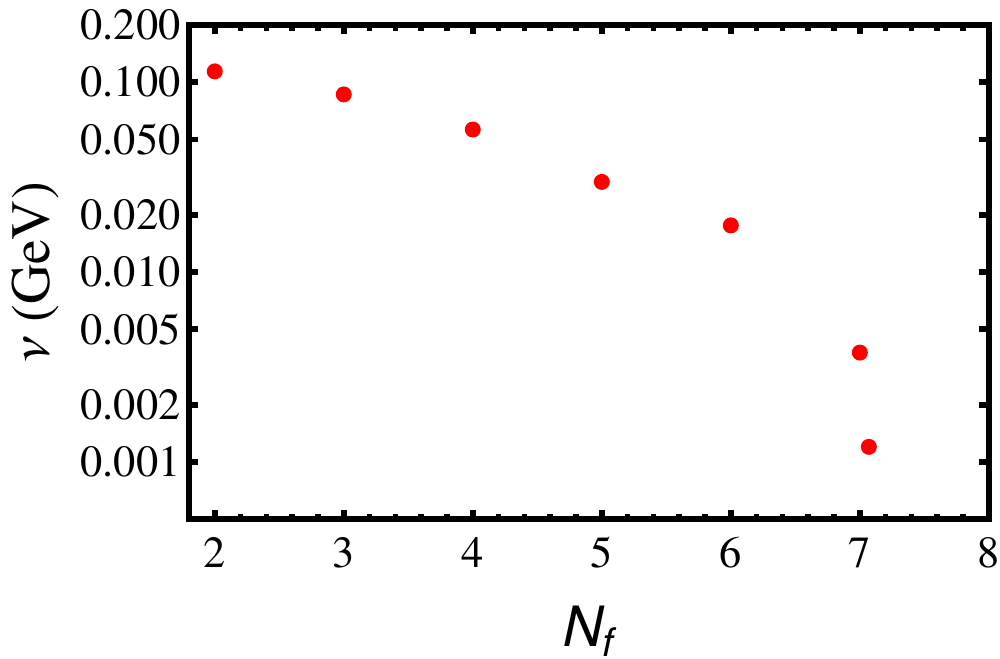} 
\end{tabular}
\end{center}
\caption{(Left) Spatially averaged Euclidean space 2-point Schwinger
function $\Delta(t)$ develops oscillations for large times which
corresponds to the non-existence of asymptotically stable free
quark states. For sufficiently large values of $N_f$, the first
minimum of these oscillations is pushed all the way to infinity,
thus ensuring the existence of a pole on the time-like axis, a
property of free particle propagators. 
(Right) The order parameter for confinement $\nu (N_f) = 1 /
\tau_1(N_f)$, where $\tau_1(N_f)$ is the location of the first
zero of Eq.~(2.12) 
Comparison with Fig.~4 
suggests that quarks get deconfined when chiral symmetry is restored.} 
\label{oscillations}
\end{figure}

Once the quark mass function is available for varying light quark
flavors (see Fig.~\ref{fig:quarkmass} for the linear case), one can investigate any of the interrelated order
parameters, namely, the Euclidean pole mass defined as $m_{\rm
dyn}^2 + M^2(p^2= m_{\rm dyn}^2)= 0$, the quark-antiquark
condensate which is obtained from the trace of the quark
propagator or the pion leptonic decay constant $f_{\pi}$ defined
through the Pagel-Stokar equation~\cite{Pagel:1979}, or through
considering the residue at the pion pole of the meson propagator.
Each of these quantities involves the quark wave-function
renormalization, the mass function and/or its derivatives and is
hence calculable from the solution for the full quark propagator.
Moreover, these order parameters can help to locate the critical
number of flavors above which chiral symmetry is restored. 

We investigate these three order parameters and choose to 
present here the Euclidean pole mass of the quark in
Fig.~\ref{fig:polemass} for the linear (exponential) case and show that, 
at a critical value of about $N_f^c \approx 7.1$ ($N_f^c \approx 9.4$), 
chiral symmetry appears restored. The phase transition appears second order, described by 
the following mean field behavior (solid lines in Fig.~\ref{fig:polemass})~:
 \bea
 m_{\rm dyn} \thicksim \sqrt{N_f^{c_2} - N_f} \;. \label{scaling}
 \eea
It has been established that confinement is related to the
analytic properties of QCD Schwinger functions which are the
Euclidean space Green functions, namely, propagators and vertices.
One deduces from the reconstruction
theorem~\cite{Reconstruction:1980s} that the only Schwinger
functions which can be associated with expectation values in the
Hilbert space of observables; namely, the set of measurable
expectation values, are those that satisfy the axiom of reflection
positivity. When that happens, the real-axis mass-pole splits,
moving into pairs of complex conjugate singularities. No
mass-shell can be associated with a particle whose propagator
exhibits such singularity structure. We  define the following
Schwinger function:
 \bea
 \Delta(t) &=& \int d^3x \int \frac{d^4p}{(2 \pi)^4} \;
 {\rm e}^{i(p_4t+ {\bf{p}} \cdot {\bf x} ) } \sigma_s(p^2) \ ,
 \label{Deltat}
 \eea
 to study the analytic properties of the quark propagator; where
 $\sigma_s(p^2)$ is the scalar term for the quark propagator in
 Eq.~(\ref{eq:Sm1}), that can be written in terms of the quark wavefunction renormalization and mass function as $Z(p^2,\mu^2)
 M(p^2)/(p^2+M(p^2))$. One can show that if there is a stable asymptotic state
 associated with this propagator, with a mass $m$, then
\bea
 \Delta(t) \thicksim {\rm e}^{-mt} \ , 
\eea
 whereas two complex conjugate mass-like singularities, with complex masses
 $\mu = a \pm i b$ lead to an oscillating behavior of the sort
\bea 
 \Delta(t) \thicksim {\rm e}^{-at} {\rm cos}(b t + \delta)
\eea
 for large $t$,~\cite{Maris:1995}.
 Fig.~\ref{oscillations} analyzes this function for varying
 $N_f$, in the linear extrapolation case. The existence of oscillations clearly demonstrates that
 the quarks correspond to a confined excitation for small $N_f$. With increasing
 $N_f$, the onslaught of oscillations moves towards higher values
 of $t$ and eventually never takes place above a critical $N_f$
 when quarks deconfine and correspond to a stable asymptotic
 state.
As an order parameter of confinement, we therefore employ $\nu
(N_f) = 1 / \tau_1(N_f)$, where $\tau_1(N_f)$ is the location of
the first singularity,~\cite{Sanchez:2009}. The first oscillation is pushed to 
infinity when confinement is lost. It is notable that when the dynamically
generated mass approaches zero, $\nu (N_f)$ diminishes rapidly 
(see right plot of~Fig.~\ref{oscillations}). This highlights the intimate
connection between chiral symmetry restoration and deconfinement.
In fact, within our numerical accuracy, $N_f^c$ is found 
to be the same for both the transitions.

The results with the exponential and linear flavor extrapolations are
qualitatively the same, leading to identical conclusions. They only
quantitatively differ by the critical flavor numbers, although 
both are pretty much in the same ballpark: $N_f^c\simeq 7.1$ and 
$N_f^c\simeq 9.4$. Note that both the parameterizations, so far apart as to
have an infinitely massive gluon at $N_f \approx 12$ or $N_f
\Rightarrow \infty$, restore chiral symmetry and trigger
deconfinement at so similar value of light quark flavors.

\section{Conclusions}

We have benefited from the latest lattice result for the quark flavor dependence of the gluon 
propagator in the infrared, as well as from a RGE-grounded nonperturbative model for this IR 
gluon propagator, in order to perform a Poincare-covariant SDE analysis for the quark propagator. 
This provided with an efficacious model for dilution of the gluon-gluon interaction with 
increasing number of light quarks and we applied it to study the evolution with the light-flavors number 
of the quark chiral behaviour. A nonperturbative picture for the chiral symmetry restoration mechanism 
emerges thus from this analysis. The quantitative analysis, following this approach, hints towards chiral symmetry 
restoration in QCD when the number of light quark flavors
exceeds a critical value of $N_f^{c_2} \thickapprox 8.2 \pm 1.2$.
This is in perfect agreement with the state-of-the-art for the direct 
lattice investigations on the chiral symmetry restoration in 
QCD~\cite{Iwasaki:2012,Aoki:2013} and supports that the model presented 
here for the chiral restoration mechanism is properly capturing most of the relevant 
physics for the problem we deal with.

Furthermore, we have also studied an order parameter for the QCD confinement-deconfinement transition, which is based on 
the analytic properties of the quark two-point Schwinger function, and located numerically the critical point at the same 
number of flavors as chiral symmetry restoration took place. That chiral symmetry appears restored 
when quarks get deconfined, within the approach we followed, highlights the intimate connection betweem both 
fundamental phenomena of QCD. This is a main result of this work.

\bigskip

\noindent {\bf Acknowledgments} We acknowledge D. Schaich for a fruitful communication. 
This work was supported  by the grants: CIC, UMICH, Mexico, 4.10 and 4.22, CONACyT (Mexico)
82230 and 128534, and MINECO (Spain) research project
FPA2011-23781.


\begin{thebibliography}{99}

\bibitem{Bashir:2013zha}
  A.~Bashir, A.~Raya and J.~Rodriguez-Quintero,
Phys. Rev. D {\bf 88}, 054003 (2013);  arXiv:1302.5829 [hep-ph].


\bibitem{Weinberg:1979} S.~Weinberg, Phys. Rev. D {\bf 13}, 974 (1976);  S.~Weinberg,
Phys. Rev. D {\bf 19}, 1277 (1979); L.~Susskind, Phys. Rev. D {\bf
20}, 2619 (1979).

\bibitem{Holdom:1985} B.~Holdom, Phys. Lett. B {\bf 150}, 301 (1985);
 K.~Yamawaki, M.~Bando and K.~Matumoto, Phys. Rev. Lett. {\bf 56},
 1335 (1986); T.~W.~Appelquist, D.~Karabali and L.~C.~R.~Wijewardhana, Phys. Rev. Lett. {\bf 57}, 957 (1986).

\bibitem{Wilczek:1973} D.~J.~Gross and F.~Wilczek, Phy. Rev. Lett. {\bf 30},
1343 (1973); H.~D. Politzer, Phy. Rev. Lett. {\bf 30}, 1346
(1973).

\bibitem{Appelquist:2009} T.~Appelquist {\em et. al.}, Phys. Rev. Lett. {\bf 104}, 071601 (2010);
 T.~Appelquist, G.~T.~Fleming and E.~T.~Neil, Phys. Rev. D {\bf 79},
076010 (2009);
 Z.~Fodor {\em et. al.}, Phys. Lett. B {\bf 681}, 353 (2009);
 K.-I.~Nagai {\em et. al.}, Phys. Rev. D {\bf 80}, 074508 (2009);
 L.~Del Debbio {\em et. al.}, Phys. Rev. D {\bf 82}, 014510 (2010);
 A.~Hasenfratz, Phys. Rev. D {\bf 82}, 014506 (2010).

\bibitem{Iwasaki:2012} M.~Hayakawa, K.-I.~Ishikawa, Y.~Osaki, S.~Takeda, S.~Uno,
N.~Yamada, Phys. Rev. D {\bf 83} 074509 (2011); {\em "Approaching
Conformality with Ten Flavors"}, T.~Appelquist, R.~C.~Brower,
M.~I.~Buchoff, M.~Cheng, S.~D.~Cohen, G.~T.~Fleming, J.~Kiskis,
M.~Lin, H.~Na, E.~T.~Neil, J.~C.~Osborn, C.~Rebbi, D.~Schaich,
C.~Schroeder, G.~Voronov, P.~Vranas, arXiv:1204.6000 [hep-ph]
(2012) ; Y.~Iwasaki, {\em "Conformal Window and Correlation
Functions in Lattice Conformal QCD"}, e-Print: arXiv:1212.4343
[hep-lat] (2012); {\em "Scale-dependent Mass
Anomalous Dimension from Dirac Eigenmodes"}, A. Cheng, A.
Hasenfratz, G. Petropoulos, D. Schaich, arXiv:1301.1355 [hep-lat]
(2013). 

\bibitem{Aoki:2013}
Y. Aoki, T. Aoyama, M. Kurachi, T. Maskawa, K-I Nagai, H. Ohki, A. Shibata, K.
Yamawaki, T. Yamazaki;  {\em "Walking Signals in Nf=8 QCD on the Lattice"}, 
arXiv:1302.6859 [hep-lat] (2013).


 \bibitem{SD:1949} F.~Dyson, Phys. Rev. {\bf 75}, 1736 (1949); J.~S.~Schwinger, Proc. Nat.
 Acad. Sci. {\bf 37}, 452 (1951); J.~S.~Schwinger, Proc. Nat. Acad.
 Sci. {\bf 37}, 455 (1951).

\bibitem{Vertex:All} J.~S.~Ball and T-W.~Chiu, Phys. Rev. D {\bf 22}, 2542
(1980); D.~C.~Curtis and M.~R.~Pennington, Phys. Rev. D {\bf 42},
4165 (1990); Z.~Dong, H.~J.~Munczek and C.~D.~Roberts, Phys. Lett.
B {\bf 33}, 536 (1994); A.~Bashir and M.~R.~Pennington, Phys. Rev.
D {\bf 50}, 7679 (1994); A.~Kizilersu and M.~R.~Pennington, Phys.
Rev. D {\bf 79} 125020 (2009).

\bibitem{Sanchez:2011} A.~Bashir, A.~Raya, S.~S\'anchez-Madrigal, Phys. Rev. D {\bf 84}, 036013 (2011).

\bibitem{Bermudez:2012} A.~Bashir, R.~Bermudez, L.~Chang,
C.~D.~Roberts, Phys. Rev. C {\bf 85}, 045205 (2012).

\bibitem{Maris:1998} P.~Maris, C.~D.~Roberts and P.~C.~Tandy, Phys. Lett. B {\bf 420}, 267
(1998); P.~Maris and C.~D.~Roberts, Phys. Rev. C {\bf 58} 3659
(1998).

\bibitem{Review:2012} A.~Bashir, L.~Chang, I.~C.~Clot, B.~El-Bennich,
Y-X.~Liu, C.~D.~Roberts, P.~C.~Tandy, Commun. Theor. Phys. {\bf
58}, 79, (2012).


\bibitem{Gluon:2009}
  I.~L.~Bogolubsky, E.~M.~Ilgenfritz, M.~Muller-Preussker and A.~Sternbeck,
   Phys. Lett. B {\bf 676}, 69 (2009); A.~C.~Aguilar, D.~Binosi, and J.~Papavassiliou, Phys. Rev.
   D {\bf 78}, 025010 (2008); P.~Boucaud, J.~Leroy, A.~L.~Yaouanc, J.~Micheli, O.~P\'{e}ne, and
   J.~Rodr\'{\i}guez-Quintero, J. High Energy Phys. {\bf 06}, 099 (2008); D.~Dudal, J.~A.~Gracey,
   S.~P.~Sorella, N.~Vandersickel, and H.~Verschelde, Phys. Rev. D 78, 065047 (2008).

\bibitem{Baron:2010bv}
  R.~Baron, P.~Boucaud, J.~Carbonell, A.~Deuzeman, V.~Drach, F.~Farchioni, V.~Gimenez and G.~Herdoiza {\it et al.},
  JHEP {\bf 1006}, 111 (2010).

\bibitem{Baron:2011sf}
  R.~Baron {\it et al.}  [ETM Collaboration],
  PoS LATTICE {\bf 2010}, 123 (2010).

\bibitem{Blossier:2010ky}
  B.~Blossier {\it et al.}  [ETM Collaboration], Phys. Rev. D {\bf 82}, 034510 (2010).

\bibitem{Blossier:2011tf}
  B.~Blossier, P.~.Boucaud, M.~Brinet, F.~De Soto, X.~Du, M.~Gravina, V.~Morenas and O.~Pene {\it et al.},
  Phys.\ Rev.\ D {\bf 85}, 034503 (2012),

\bibitem{Frezzotti:2000nk}
  R.~Frezzotti {\it et al.}  [Alpha Collaboration],  JHEP {\bf 0108}, 058
  (2001).

\bibitem{Bogolubsky:2007ud}
  I.~L.~Bogolubsky, E.~M.~Ilgenfritz, M.~Muller-Preussker and A.~Sternbeck,
  PoS LAT {\bf 2007}, 290 (2007).

\bibitem{Ayala:2012}
  A.~Ayala, A.~Bashir, D.~Binosi, M.~Cristoforetti and J.~Rodriguez-Quintero,
  Phys. Rev. D {\bf 86}, 074512(2012).

\bibitem{Aguilar:2012rz}
  A.~C.~Aguilar, D.~Binosi and J.~Papavassiliou,
  Phys.\ Rev.\ D {\bf 86} (2012) 014032
  [arXiv:1204.3868 [hep-ph]].


\bibitem{Blossier:2012ef}
  B.~Blossier, P.~.Boucaud, M.~Brinet, F.~De Soto, X.~Du, V.~Morenas, O.~Pene and K.~Petrov {\it et al.},
  Phys.\ Rev.\ Lett.\  {\bf 108} (2012) 262002
  [arXiv:1201.5770 [hep-ph]];
  B.~Blossier {\it et al.}  [ETM Collaboration],
  arXiv:1310.3763 [hep-ph].


\bibitem{Dudal:2012zx}
D.~Dudal, O.~Oliveira and J.~Rodriguez-Quintero, Phys. Rev. D {\bf
86}, 105005 (2012).


\bibitem{Dudal:2008sp}
  D.~Dudal, J.~A.~Gracey, S.~P.~Sorella, N.~Vandersickel and H.~Verschelde,
  Phys.\ Rev.\ D {\bf 78} (2008) 065047
  [arXiv:0806.4348 [hep-th]].

\bibitem{Dudal:2010tf}
  D.~Dudal, O.~Oliveira and N.~Vandersickel,
  Phys.\ Rev.\ D {\bf 81} (2010) 074505
  [arXiv:1002.2374 [hep-lat]].

\bibitem{Boucaud:2000nd}
  P.~Boucaud, A.~Le Yaouanc, J.~P.~Leroy, J.~Micheli, O.~Pene and J.~Rodriguez-Quintero,
  Phys.\ Lett.\ B {\bf 493} (2000) 315
  [hep-ph/0008043];
  F.~V.~Gubarev and V.~I.~Zakharov,
  Phys.\ Lett.\ B {\bf 501} (2001) 28
  [hep-ph/0010096];
  K.~-I.~Kondo,
  Phys.\ Lett.\ B {\bf 514} (2001) 335
  [hep-th/0105299];
  H.~Verschelde, K.~Knecht, K.~Van Acoleyen and M.~Vanderkelen,
  Phys.\ Lett.\ B {\bf 516} (2001) 307
  [hep-th/0105018];
  D.~Dudal, H.~Verschelde and S.~P.~Sorella,
  Phys.\ Lett.\ B {\bf 555} (2003) 126
  [hep-th/0212182];
  E.~Ruiz Arriola, P.~O.~Bowman and W.~Broniowski,
  Phys.\ Rev.\ D {\bf 70} (2004) 097505
  [hep-ph/0408309].

\bibitem{Boucaud:2013bga}
  P.~.Boucaud, M.~Brinet, F.~De Soto, V.~Mor\`enas, O.~P\`ene, K.~Petrov and J.~Rodr\'{\i}guez-Quintero,
  arXiv:1312.1514 [hep-lat].

\bibitem{Pagel:1979} H.~Pagels and S.~Stokar, Phys. Rev. D {\bf 20} 2947 (1979).

\bibitem{Reconstruction:1980s} R.~F.~Streater and A.~S.~Wightman, {\em "PCT, spin and statistics, and
all that,"} (1989); J.~Glimm and A.~Jaffee, {\em "Quantum Physics.
A Functional Point of View,"} Springer-Verlag, New York(1981).

\bibitem{Maris:1995} P.~Maris, Phys. Rev. D {\bf 52} 6087 (1995).

\bibitem{Sanchez:2009} A.~Bashir, A.~ Raya, S.~S\'anchez-Madrigal, C.D.~Roberts, Few Body
Sys. {\bf 46}, 229 (2009).


\end{thebibliography}
\end{document}